\documentclass[floatfix]{revtex4}
\usepackage[dvips]{epsfig,graphics}
\usepackage{graphicx}
\usepackage{amsmath}
\usepackage{amssymb}
\usepackage{amsfonts}

\begin{document}

\title{Can the LHC rule out the MSSM ?}
\author{S.~S.~AbdusSalam}
\affiliation{Abdus Salam ICTP, Strada Costiera 11, I-34014 Trieste, Italia}

\begin{abstract}
If supersymmetry (SUSY) exists in nature and is a solution to the
hierarchy problem then it should be detectable at the
TeV energy scale which the large hadron collider (LHC) is now
exploring. One of the main goals of the LHC is the 
discovery or exclusion of the R-parity conserving minimal
supersymmetric standard model 
(MSSM). So far, the SUSY search results are presented in the context
of the constrained MSSM and other specific simplified SUSY models.
 A model-independent analysis necessarily  relies on the
 trigger-system of the LHC detectors.
  By using the posterior samples of a
20-parameter MSSM, the phenomenological MSSM, from a fit to indirect
collider and cosmological data we find 
that there is a significant volume in the MSSM parameter space that
would escape the standard trigger-systems of the detectors. As such, in the
absence of discovery in the current and future LHC runs, it would be
difficult if not impossible to exclude the MSSM unless some dedicated
and special triggers are commissioned or a Higgs 
boson with mass as predicted by the supersymmetric models is not found.  
\end{abstract}

\maketitle

\paragraph*{Introduction.}
The current understanding of nature, in the form of experimentally
tested knowledge about 
the fundamentals of particle physics, have major limitations 
that include the ignorance about what constitutes cold dark matter
\cite{Bertone:2004pz} 
(CDM) and about the mechanism for explaining particles masses and the 
wide hierarchies between them.  The
Higgs mechanism could explain the source for particle masses
but suffers the so-called hierarchy problem since the mass of 
the Higgs particle itself is not stable to radiative
corrections~\cite{Martin:1997ns}. 

A solution for the 
hierarchy problem is feasible with supersymmetry (SUSY) by which
the Higgs boson mass become stable to radiative corrections.
SUSY predicts that for each standard model (SM) particle there 
should be a corresponding equal-mass sparticle with a half-unit
spin difference. 
The fact that no sparticle has been
observed to date indicates that the symmetry, if it exists,
is broken and the sparticles are much heavier than their corresponding
SM particles. It is expected that sparticles would be produced at high
energy colliders. For instance, from a proton-proton collision
strongly interacting squarks($\tilde q$) and gluinos($\tilde g$) would be pair-produced which
would subsequently decay into SM particles to form jets, leptons and
photons. With  
R-parity conservation~\cite{Farrar:1978xj}, sparticle decay chains
would end with the lightest supersymmetric particle (LSP) which is
cosmologically stable and hence could account for the whole or
part of the CDM~\cite{AbdusSalam:2010qp}. It is expected that a
SUSY scenario that addresses the hierarchy-problem  would be manifest at
${\mathcal {O}}$(TeV) energy-scale. 

At particle colliders where sparticles 
could be produced,
the neutralino($\tilde \chi^0_1$) LSP would 
fly out without interacting with the collider particles
detection systems. Its experimental signature corresponds to 
final states with unbalanced total 
momentum in the transverse plane to the colliding particles beam axis. The
magnitude of the missing transverse energy ($E_T$) and the transverse
momentum, $p_T$, of the resulting jets respectively depends on the LSP
mass and its difference from the lightest squark or
gluino. Generically, SUSY events at colliders
would have large missing $E_T$ (LSP mass) 
and large jets $p_T$ (related to the energy carried by quarks from
heavy squark/gluino decay to neutralino.) 
However, this is not necessary always the case as it will be shown in
the forthcoming paragraphs. SUSY scenarios with non-standard
characteristics should not be neglected in strategic search  studies.

The LHC~\cite{Evans:2008zzb} is a proton-proton collider designed to
reach up to 14 TeV centre-of-mass energy in search for the origin of
particles masses and for new physics beyond the SM (BSM). 
During the year 2010 runs, the LHC general purpose detectors, ATLAS and
CMS, have collected about 35 pb$^{-1}$ data at 7 TeV. No significant
excess to the SM expectation has been 
found in their SUSY search analyses. The results are 
reported in the form of limits on the parameter space of some chosen model 
and on non-SM cross sections for the chosen SUSY search
channels. 

In this article we emphasis that the LHC
results are detectors' trigger-system dependent.  
SUSY final states with characteristics that cannot be triggered by the
detectors would be lost forever as "uninteresting'' event(s). Events
that end up with very soft jets would be rejected as 
 QCD background (non-SUSY) since the triggers are optimised to minimise
the recording of background events. For example,
the non-SM cross section limits presented in Ref.~\cite{daCosta:2011qk}
would necessary not apply for the SUSY scenario where the resulting
jets and missing energy would not get triggered or would escape
the imposed kinematic cuts. Therefore, in the
absence of SUSY discovery in the current and future LHC 
runs, it would be difficult, if not impossible, to rule out the
MSSM. We give explicit examples of such SUSY points from the
20-parameter phenomenological MSSM (pMSSM)  
fits' (to indirect collider and cosmological data) posterior
samples~\cite{AbdusSalam:2009qd}, 
which are not possible to obtain from the constrained MSSM
(CMSSM/mSUGRA) parameter space, that would escape the LHC detectors'
trigger-systems  
and perform a generator-level analyses on simulated SUSY events to
illustrate the nature of the resulting jets and leptons $p_T$ and of
events missing $E_T$ that could be obtained.

\paragraph*{The pMSSM.}
The pMSSM~\cite{Haber1998,Djouadi:1998di,Profumo:2004at,
  AbdusSalam:2008uv,AbdusSalam:2009qd,Berger:2008cq}  
is set up such that the physics behind
SUSY breaking, mediation mechanism and renormalisation group (RG) running of the parameters
from the SUSY-breaking scale is not relevant for, and hence decoupled
from, the SUSY-breaking terms parameterisation procedure. The soft
SUSY-breaking parameters (20 and a sign choice) are derived from the
parent 105-parameters MSSM by imposing phenomenological constraints to  
suppress CP-violation sources and constraints from the observed 
absence of FCNC: 
\begin{align}
M_{1,2,3};\; m^{3rd \, gen}_{\tilde{f}_{Q,U,D,L,E}},\;
m^{1st/2nd \, gen}_{\tilde{f}_{Q,U,D,L,E}}; A_{t,b,\tau,\mu=e}, m^2_{H_{u,d}},
\tan \beta 
\nonumber
\end{align}
where the first 3 parameters, $M_1$, $M_2$ and $M_3$ are the gaugino
mass parameters. The next 10 parameters represent 3rd and degenerate
1st and 2nd generation sfermino masses. $A_{t,b,\tau,\mu=e}$ represent
the trilinear scalar couplings. An alternative choice of parameters
would be to replace the Higgs doublets mass terms $m^2_{H_u}$ and $
m^2_{H_d}$ with the CP-odd neutral Higgs mass parameter, $M_A$ and the
Higgs doublets mixing parameter, $\mu$. 
As illustrated in
Ref.~\cite{AbdusSalam:2008uv,AbdusSalam:2009qd} the 
technology to simultaneously vary and explore the parameters for
making a statistically convergent global fit is well 
within reach. However in the absence of direct SUSY data the pMSSM
parameter space is weakly constrained and most, but not all, posterior
inference would be necessarily prior dependent. 
A Bayesian comparison/selection analyses between 
GUT-scaled models for SUSY-breaking is performed in
Ref.~\cite{AbdusSalam:2009tr} and pMSSM discovery prospects were
studied in Ref.~\cite{prospe}.

\begin{figure}[th]
 \begin{tabular}{ll}
  \centering\includegraphics[width=.45\textwidth]{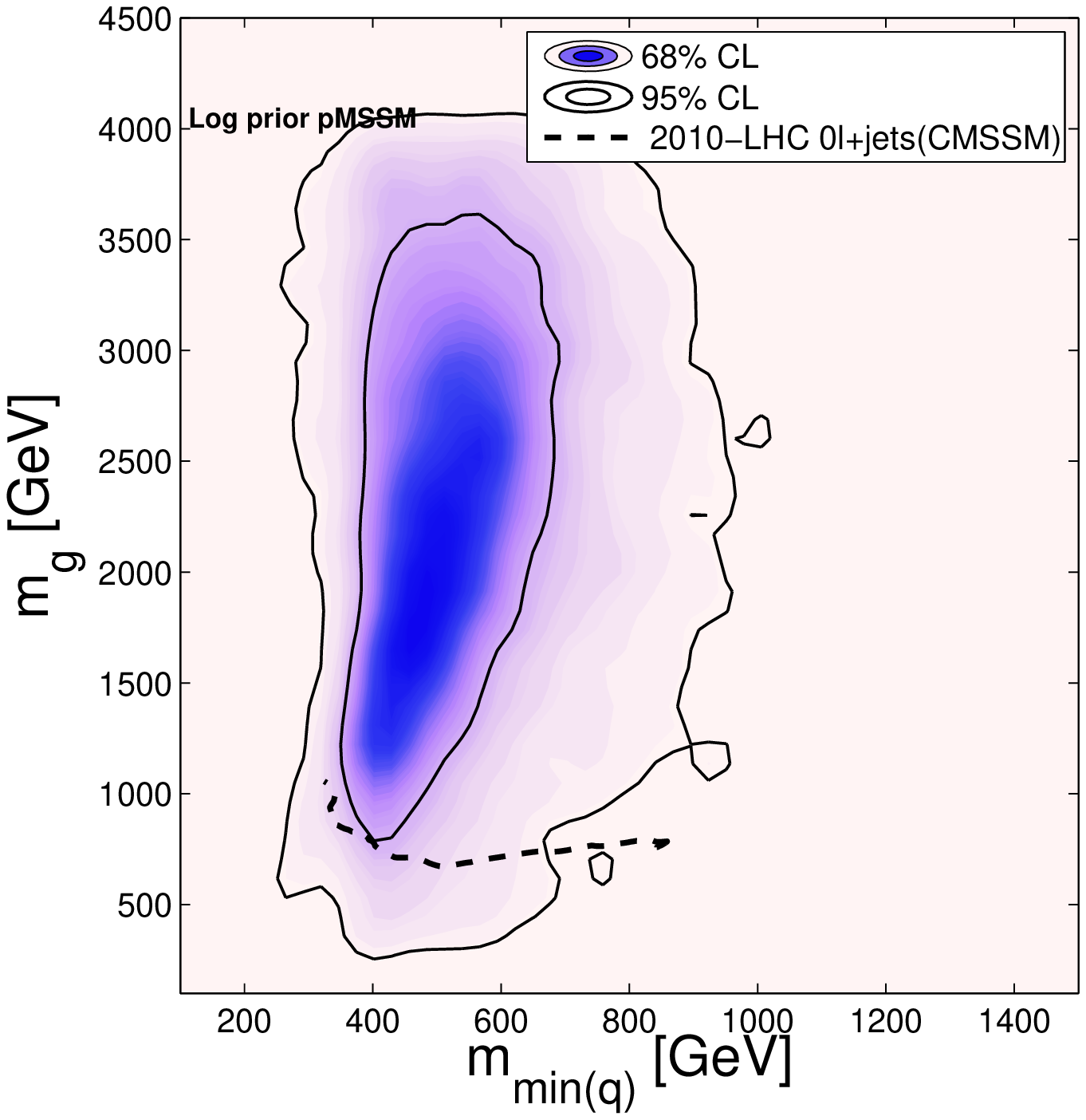}&
  \centering\includegraphics[width=.45\textwidth]{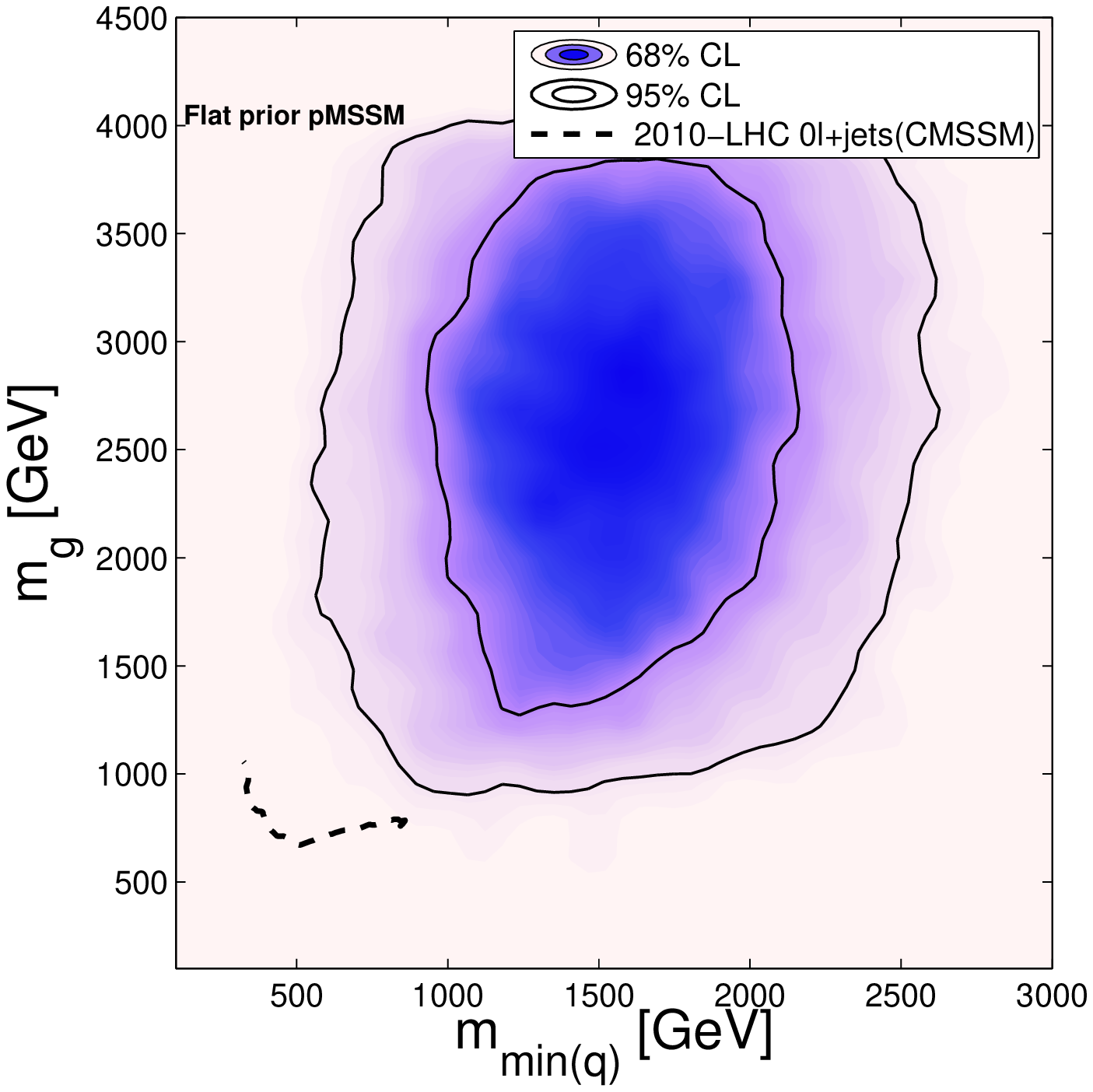}
 \end{tabular}
 \caption{pMSSM posterior probability map for gluino and lightest
   squark masses based on fits to indirect collider and cosmological 
   data~\cite{AbdusSalam:2008uv,AbdusSalam:2009qd}. The inner and
   outer contours represent $68\%$ and $95\%$ probability regions
   respectively. For reference purpose a contour line approximating
   the LHC limits~\cite{daCosta:2011qk,CMS-PAS-SUS-10-005} for
   CMSSM/mSUGRA is also shown (dotted line). Points below the contour
   are excluded at $95 \%$ CL for the CMSSM/mSUGRA models but not
   necessarily so for the pMSSM.} 
 \label{fig.4pmssm} 
\end{figure}

The squark and gluino mass posterior probability distributions from
the pMSSM fits to indirect collider and cosmological data are shown in
Fig.~\ref{fig.4pmssm}. For both log and flat prior cases the region
with $m_{\tilde g} > m_{\tilde q}$ 
is preferred. 
A contour approximating the experimental  
exclusion limits~\cite{daCosta:2011qk,CMS-PAS-SUS-10-005} on the
CMSSM/mSUGRA squark-gluino plane is also placed on top of the pMSSM
posterior distribution for reference purpose. 

There are various SUSY spectrum topologies in the pMSSM posterior
points which are difficult to obtained from constrained versions of
the MSSM. In particular
 an 
approximately neutralino degenerate squarks in the CMSSM/mSUGRA
parameter space is difficult to obtain. That is mainly due to the nature
of the RG running boundary conditions at the GUT-scale: the ratio
$M_1:M_2:M_3 = 1:2:6$ is fixed at the electroweak (EW) scale and to
leading order,  
$M_i/g_i$ do not RG run~\cite{Martin:1997ns}. Here $g_i, i=1,2,3$
represents the electromagnetic, weak or strong interaction couplings
respectively; $M_1$, $M_2$, and $M_3$ are the gaugino soft mass
parameters at the EW scale. For a predominantly bino  
neutralino 
the
gluino-neutralino mass ratio is fixed. Next, the gluino mass scales as
$m_{\tilde g} \sim M_3$ correct to subleading corrections after RG
running from the GUT scale to the EW scale. On the
other hand, squark mass $m^2_{\tilde q} \sim m_0^2 + K_3 + \ldots$,
where $K_3 \sim 6.0 m_{1/2}^2$, where $m_{1/2}$ is the common gaugino
soft mass term at the GUT scale. 
Squarks would therefore be mostly 
heavier than the gluino, so obtaining a compact spectrum $m_{\tilde g}
>> m_{\tilde q} \gtrsim m_{\tilde \chi^0_1}$, such as the one shown in
Fig.~\ref{tab.spec}(b), would be very rare/impossible. 

\begin{figure}[th]
  \begin{tabular}{cc}
  (a) & (b) \\
  \begin{minipage}[t]{8.5cm}
  \centering\includegraphics[width=.95\textwidth,height=.95\textwidth]{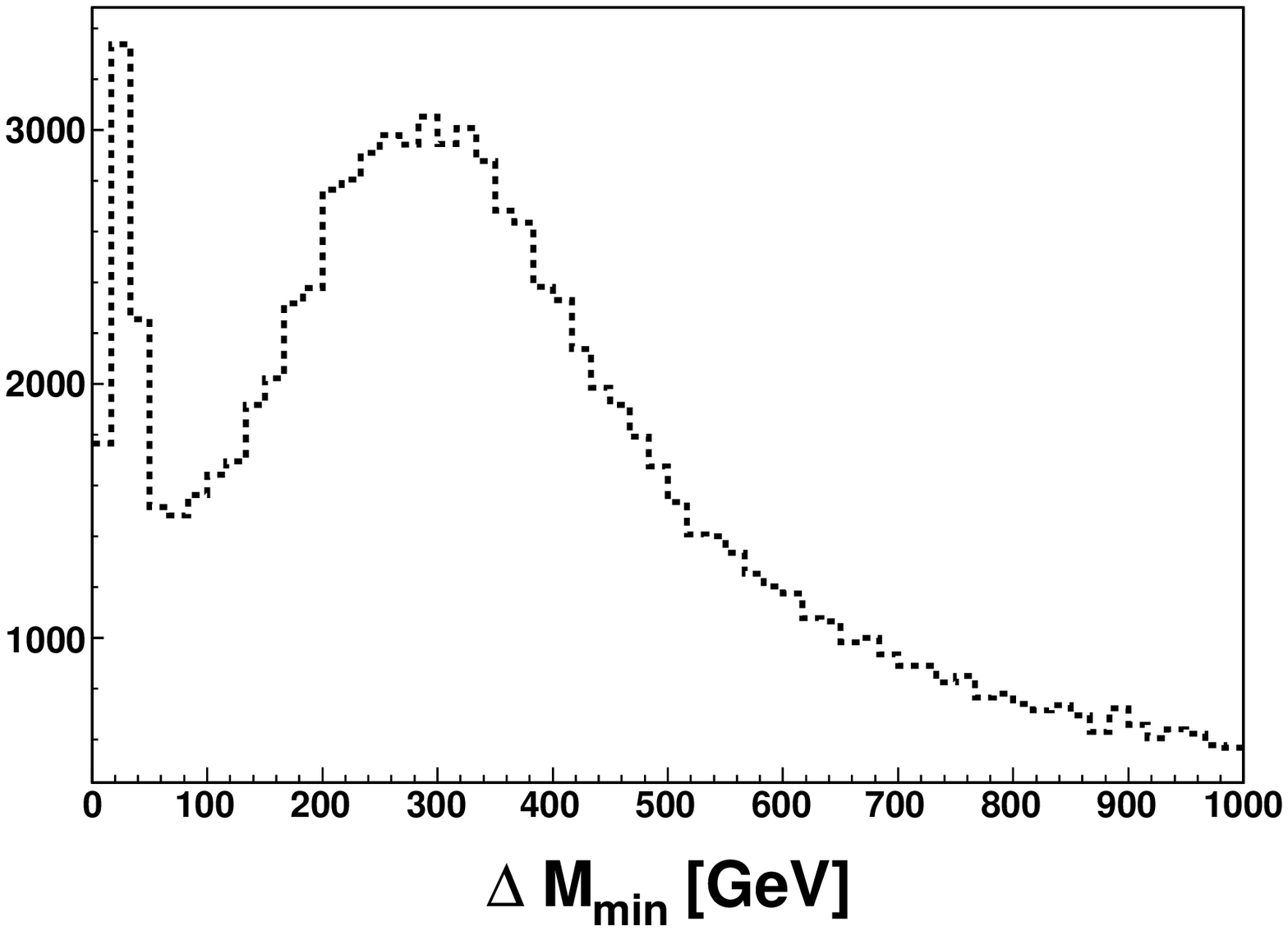}
 \end{minipage}
  &
  \begin{minipage}[t]{8.5cm}
  \centering\includegraphics[width=.95\textwidth,angle=90]{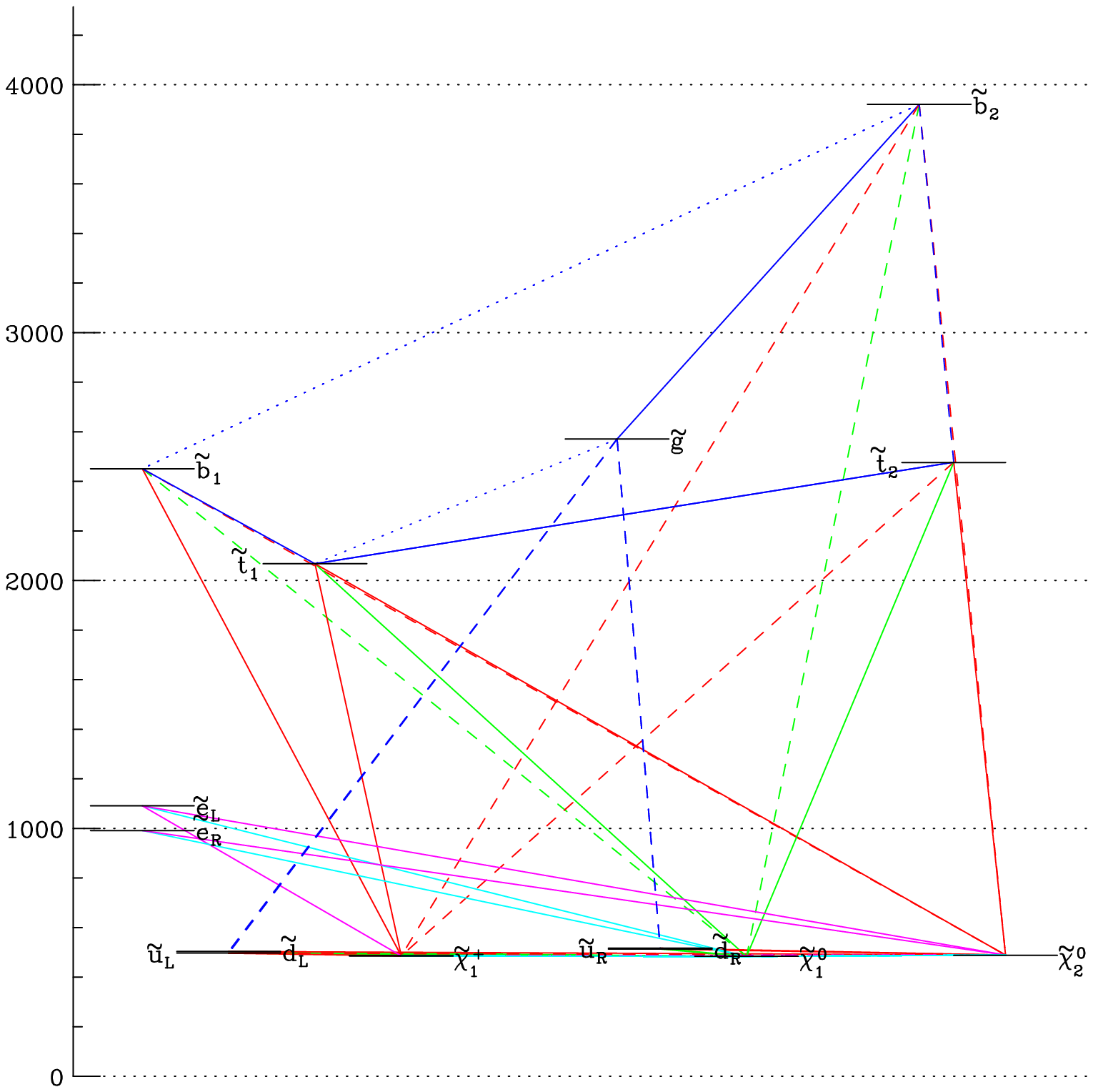} 
  \end{minipage}
  \end{tabular}
 \caption{{\bf (a)} The distribution of $\Delta M_{min}= m_{min(\tilde
     q)} -  m_{\tilde \chi^0_1}$  from 116931 pMSSM posterior
   points, about $17\%$ have $\Delta M_{min} << m_t$ centered
   around 20 GeV. {\bf (b)} Example pMSSM spectrum with
   $\Delta M_{min}= m_{min(\tilde q^{1st/2nd \, gen})} -  m_{\tilde
     \chi^0_1} << m_t$, $\Delta M_{max} = m_{max(\tilde q^{1st/2nd \,
       gen})} - m_{\tilde \chi^0_1} << m_t$ and $m_{\tilde q^{1st/2nd
       \, gen}} \sim m_{\tilde{\chi}_1^0} << m_{\tilde g, \tilde
     q^{3rd \, gen}} > 2\textrm{ TeV}$. Solid, dashed and dotted lines
   approximately represent sparticle decays 
   with branching ratios respectively greater or equal to $0.1$,
   $0.01$ and $0.001$ obtained from
   Herwig~v6.510~\cite{Corcella:2002jc} interfaced with Isajet
   v7.72. The sparticle and Higgs boson masses are given in
   Tab.~\ref{tab.spec} (spectrum Z.)}   
 \label{deltM}
\end{figure}

For the pMSSM, the scenario is different. All SUSY-breaking parameters
are freely varied at the EW scale so constraints on the parameters
from RG running is minimal. The pMSSM fits prefer $m_{\tilde g} > m_{\tilde
  q_{min}}$ as shown in Fig.~\ref{fig.4pmssm}. Moreover there is a
significant number of pMSSM 
parameter points where the mass difference between the lightest squark 
and the LSP is small, compared to the top-quark mass, that would
lead to soft jets in collider (SUSY event) final states. 
About $17\%$ of the 116931 pMSSM 
posterior points considered have lightest squarks quasi-degenerate 
with the neutralino LSP. These model points are concentrated around the $\Delta
M_{min} = m_{min(\tilde q)} - m_{\tilde \chi^0_1}$ equals 10 to 25 GeV
narrow-peak region shown in Fig.~\ref{deltM}(a) and have
various squark masses and decay patterns. For instance, only 81 points
remain if both $\Delta M_{min} = m_{min(\tilde   q)^{1st/2nd \, gen}}
- m_{\tilde \chi^0_1}$ and $\Delta M_{max} = m_{max(\tilde q)^{1st/2nd
    \, gen}} - m_{\tilde \chi^0_1}$ are required to be within the
top-quark mass, $m_t$, 
and that $m_{\tilde q^{1st/2nd \, gen}} \sim m_{\tilde{\chi}_1^0} <<
m_{\tilde g, \tilde q^{3rd \, gen}} > 2\textrm{ TeV}$; see the example
point shown in Fig.~\ref{deltM}(b) whose sparticle and Higgs boson
masses are given in Tab.~\ref{tab.spec} (spectrum Z). These would be
difficult to exclude at the LHC due to the fact that SUSY events from
such a scenario would be mainly buried under the huge QCD background
and would, as a result of the trigger system design to suppress QCD
background, not be captured by the detectors.  

\paragraph*{Example pMSSM  points.}

\begin{table}[ht]
\begin{tabular}{|c| c c c c c|}
\hline  
Sparticles & A & B & C & D & Z\\
\hline
$\tilde{d}_L \sim \tilde{s}_L$ &  507  & 579 & 2516& 971 & 507\\
$\tilde{u}_L \sim \tilde{c}_L$ & 489  & 579 & 2521& 986& 497\\
$\tilde{b}_1$ &  1874  & 318  & 601& 525& 2450\\
$\tilde{t}_1$ & 982 & 1466 &  555& 498& 2068\\
$\tilde{e}_L \sim \tilde{\mu}_L$ & 142  & 642 & 1579& 584& 1091\\
$\tilde \nu_e \sim \tilde \nu_\mu$ & 1415&627 &1578 & 586& 1091\\
$\tilde{\tau}_1$ & 1778  & 926 & 1455& 697& 1130\\
$\tilde{\nu}_\tau$ & 2199  & 1226 & 2753& 1636& 1602\\
$\tilde{g}$ &  1390  & 441 & 587& 513& -2571\\
$\tilde{\chi}_1^0$ &  437  & 307 &  518& 157& 491\\
$\tilde{\chi}_2^0$ &  2153  & 864 &  3019& 179& 495\\
$\tilde{\chi}_1^{\pm}$ &  442  & 861 & 532& 163& 497\\
$\tilde{\chi}_3^0$ &  2988  & 867 &  3549& 356& 1398\\
$\tilde{\chi}_4^0$ &  2993  & 2756 &  3563& 447& 2353\\
$\tilde{\chi}_2^{\pm}$ &  2958  & 2789 & 3502& 355& 2344\\
$\tilde{d}_R \sim \tilde{s}_R$ &  2841  & 458 & 2020& 557& 514\\
$\tilde{u}_R \sim \tilde{c}_R$ & 1279  & 2423 & 1529& 298& 515\\
$\tilde{b}_2$ &  2806 & 2553  & 828& 1465& 3921\\
$\tilde{t}_2$ & 1893 & 2562 & 1001& 1616& 2477\\
$\tilde{e}_R \sim \tilde{\mu}_R$ & 2456  & 2380 & 1443& 870& 992\\
$\tilde{\tau}_2$ & 2194  & 1225 & 2748& 1630& 1599\\
\hline
$h$ &  121  & 120 & 117& 126& 121\\
$H_0$ &  3251  & 1265 & 2918& 2876& 502\\
$A_0$ &  3250  & 1265 & 2920& 2876& 502\\
$H^{\pm}$ &  3252  & 1267 & 2921& 2877&506 \\
\hline 
\end{tabular}
\caption{Sparticle spectrum and Higgs bosons for the example pMSSM
  points A, B, C, D and Z. All masses are in GeV.}
\label{tab.spec}
\end{table}

\begin{figure}[th]
  \begin{tabular}{cc}
  (a) & (b) \\
  \begin{minipage}[t]{8.5cm}                              
  \centering\includegraphics[width=.95\textwidth,angle=90]{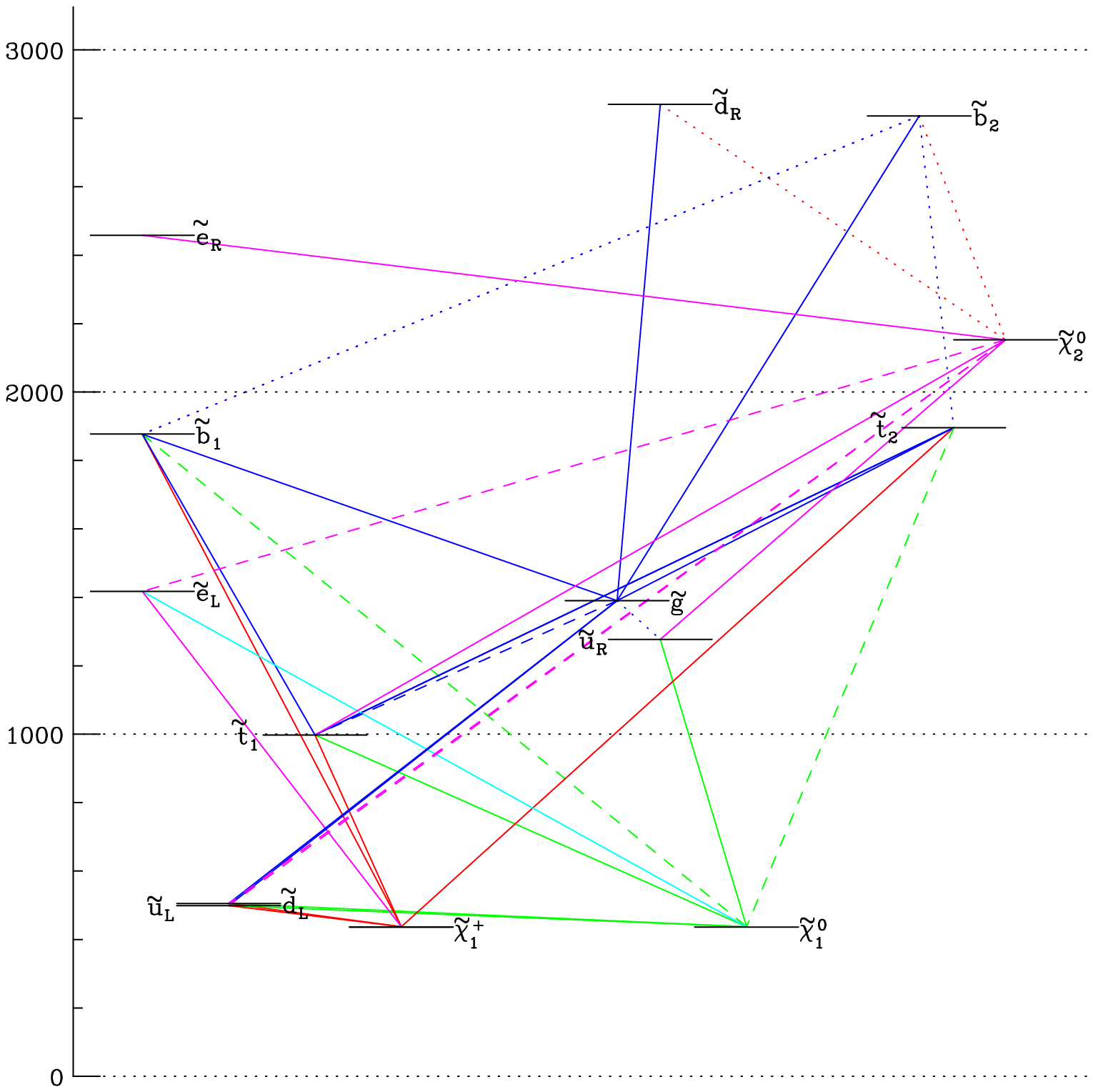}
  \end{minipage} 
                 
  &
  \begin{minipage}[t]{8.5cm}                              
  \centering\includegraphics[width=.95\textwidth,angle=90]{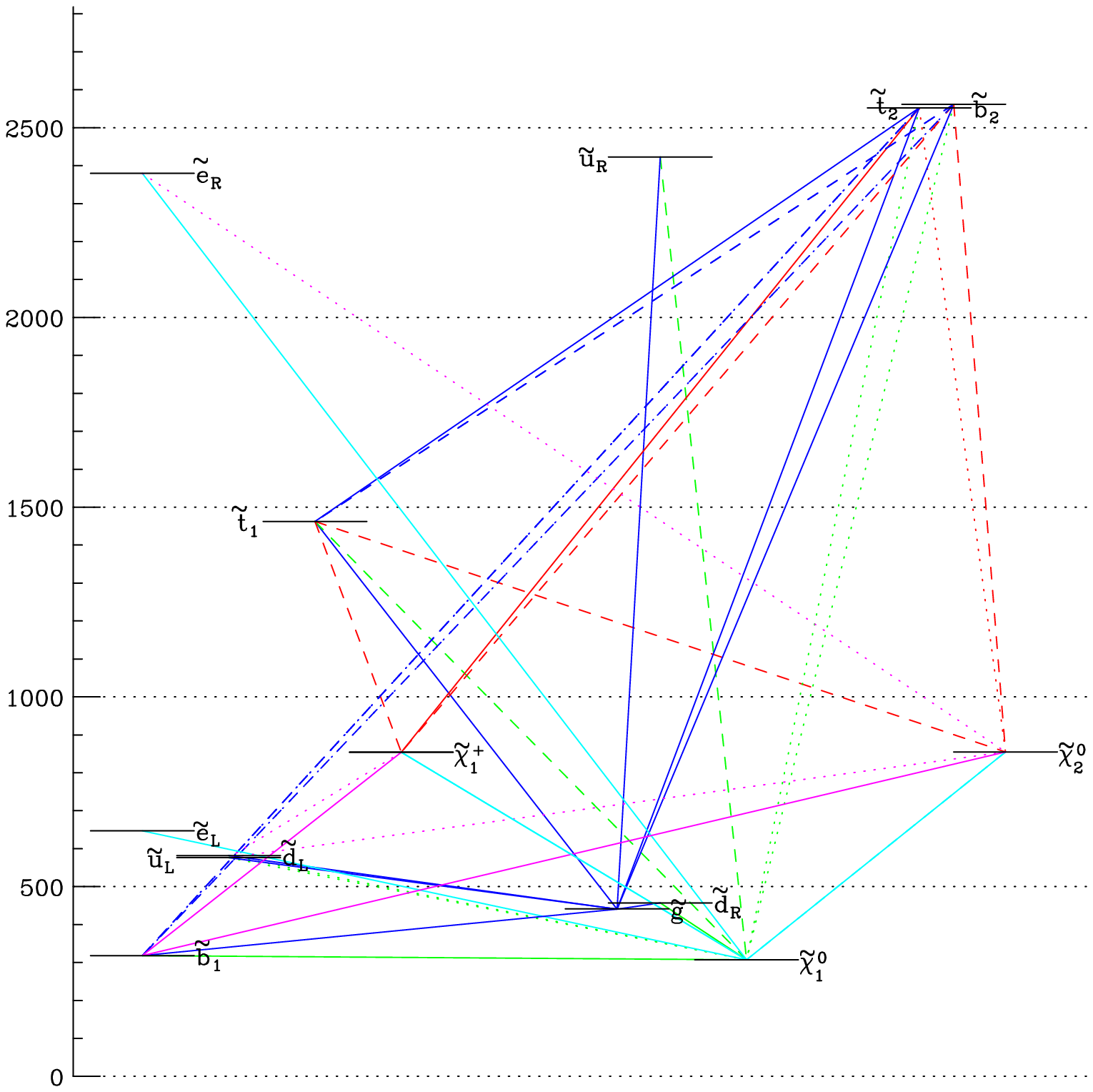}
  \end{minipage} \\  
 
  (c) & (d) \\
  \begin{minipage}[t]{8.5cm}                               
  \centering\includegraphics[width=.95\textwidth,angle=90]{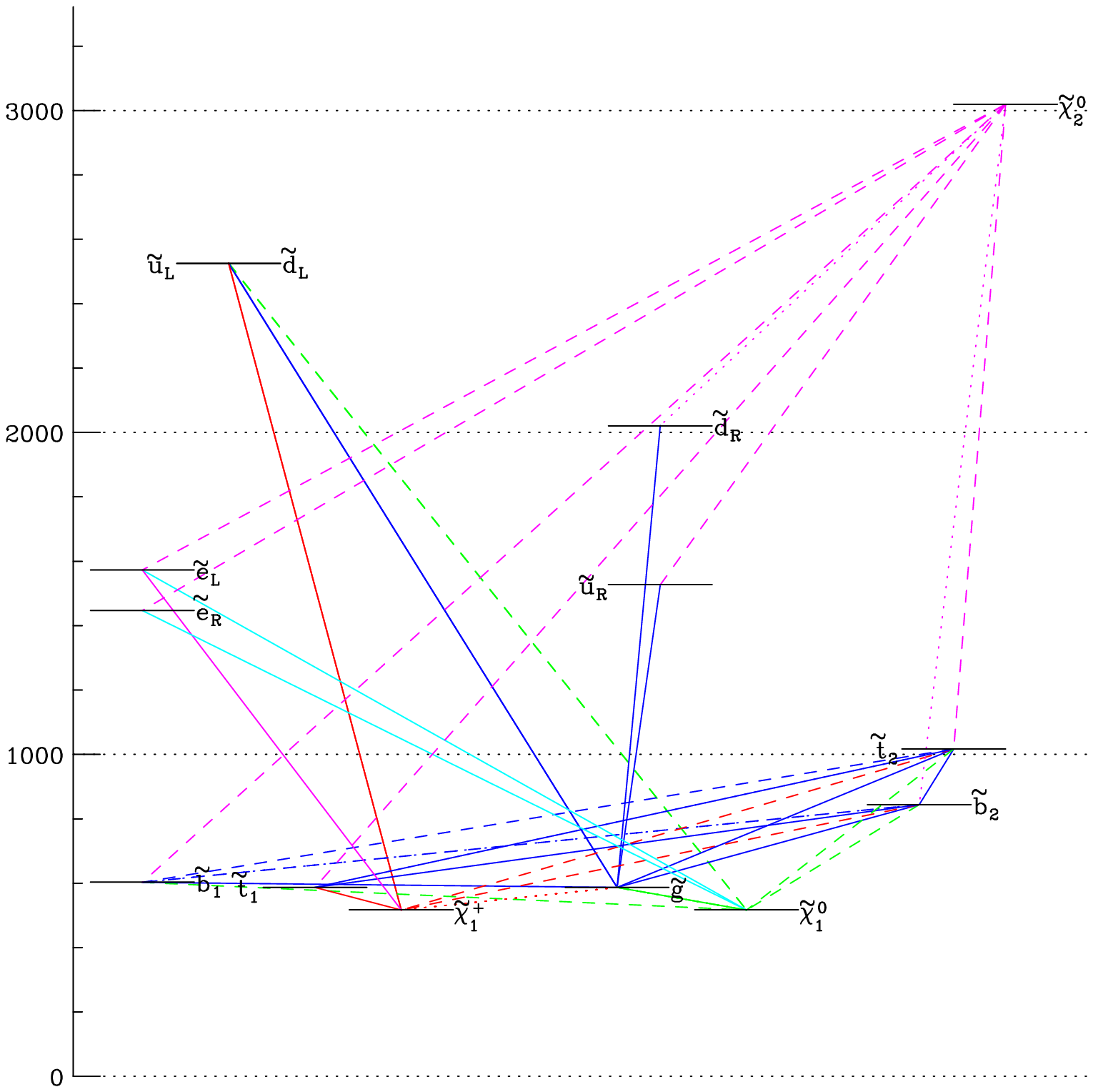}
  \end{minipage} 
  &
  \begin{minipage}[t]{8.5cm}                               
  \centering\includegraphics[width=.95\textwidth,angle=90]{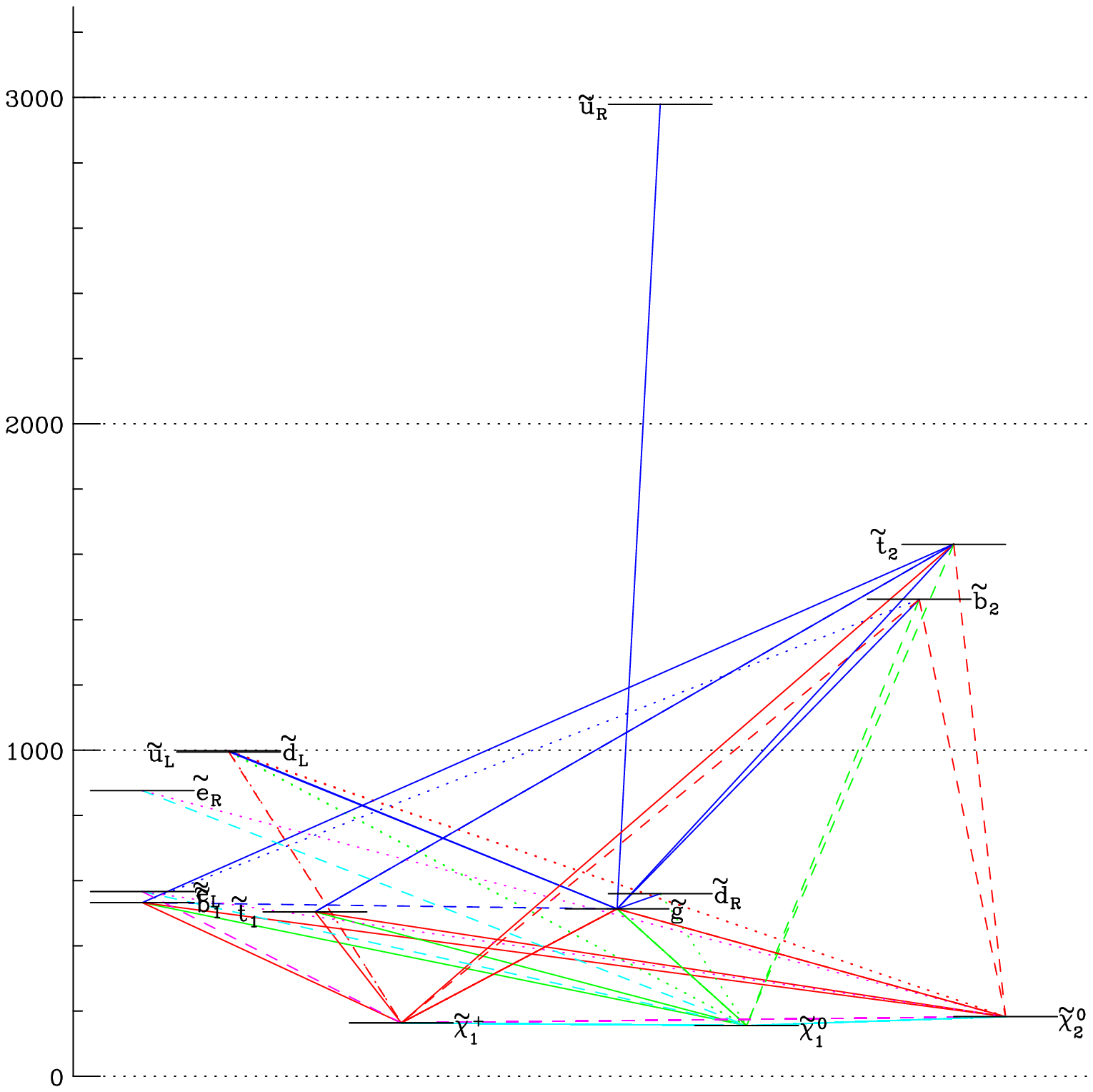}
  \end{minipage}        
  \end{tabular}
 \caption{{\bf Plots (a)} and {\bf (b)} show pictures of the example pMSSM
   spectrum A and B respectively (see Tab.~\ref{tab.br} for the mass
   magnitudes.) that have an LSP quasi-degeneracy with 1st/2nd
   generation type lightest squark and
   hence would be difficult to trigger by the LHC 2010 and
   current trigger systems. 
   The other squarks and/or gluino are much heavier than $1$ TeV so
   have relatively negligible production cross sections. The search
   for spectrum A could be via 0-lepton plus jets plus missing
   $E_T$ channel which is same for spectrum B but with additional
   possibility for b-jets. Similarly spectrum C and D, shown on {\bf
     plot (b)} and {\bf (c)} respectively, are example cases for the
   search channel involving b-jets and leptons. In
   all the plots solid, dashed and dotted lines approximately
   represent decays with branching ratios respectively greater or
   equal to $0.1$, $0.01$ and $0.001$ obtained from
   Herwig~v6.510~\cite{Corcella:2002jc} interfaced with Isajet v7.72.}  
 \label{fig.spec1}
\end{figure}

For illustration, apart from the spectrum shown in
Fig.~\ref{deltM}(b), four other example spectra labelled A, B, C, and
D, are shown in 
Fig.~\ref{fig.spec1} and  Tab.~\ref{tab.spec}. For spectrum A, SUSY
production would be dominated by the light squarks production which
would readily  decay to the neutralino and ordinary jets. Spectrum B
has an interesting 
feature that all produced squarks and gluinos would decay to the LSP
via the the lightest squark, the sbottom, so there would be
possibly for b-tagging of jets. The dominant
leading order(LO) and next-to-leading order(NLO) sparticle production
cross sections for 7 TeV proton-proton collider and selected decay
branching ratios, respectively computed using
\texttt{prospino}~\cite{Beenakker:1996ch}v2.1 and
\texttt{SUSY-HIT}~\cite{Djouadi:2006bz} packages, for the spectrum B scenario
are summarised in Tab.~\ref{tab.br}. Unlike A and B, spectrum C 
and D would have SUSY-initiated leptons in events final states.   

\begin{table}[th]
\begin{tabular}{|l c|}
\hline
Process & Branching ratio \\ 
\hline
$\tilde u_L \rightarrow \tilde g \quad u$ & 0.995 \\
$\tilde d_L \rightarrow \tilde g  \quad d$ & 0.995 \\
$\tilde d_R \rightarrow \tilde g  \quad d$ & 0.692 \\
$\tilde g \rightarrow \tilde b_1  \quad bb$ & 0.500   \\
$\tilde g \rightarrow \tilde b_1*  \quad b$ & 0.500   \\
$\tilde b_1 \rightarrow \tilde {\chi}_1^0  \quad b$ & 1.000 \\
\hline   
 $\sigma^{ss}_{LO,NLO} $  & 0.540, 0.710 \\
 $\sigma^{sg}_{LO,NLO} $  & 3.16, 5.72 \\
 $\sigma^{gg}_{LO,NLO} $  & 2.26, 4.93 \\
\hline
\end{tabular} 
\caption{Selected sparticle leading order (LO) and next-to-leading
  order (NLO) production cross sections and decay branching ratios for
  the example pMSSM spectrum B scenario for which b-tagging of jets
  can be implemented.}   
\label{tab.br}
\end{table}


\begin{figure}[th]
  \begin{tabular}{cc}
  (a) & (b) \\
  \begin{minipage}[t]{8.5cm}
  \centering\includegraphics[width=.95\textwidth,angle=90]{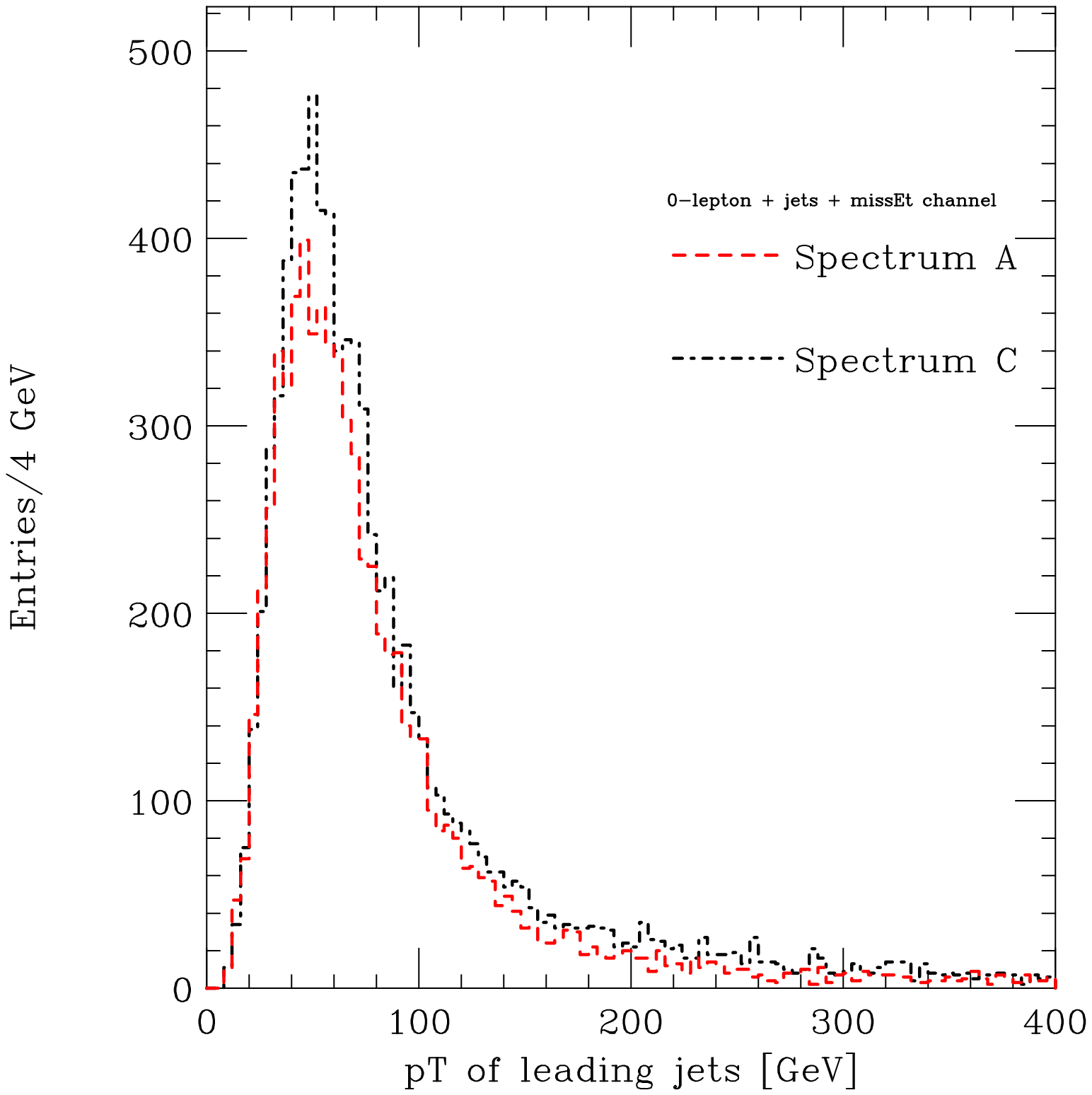}
  \end{minipage}  
  &
  \begin{minipage}[t]{8.5cm}
  \centering\includegraphics[width=.95\textwidth,angle=90]{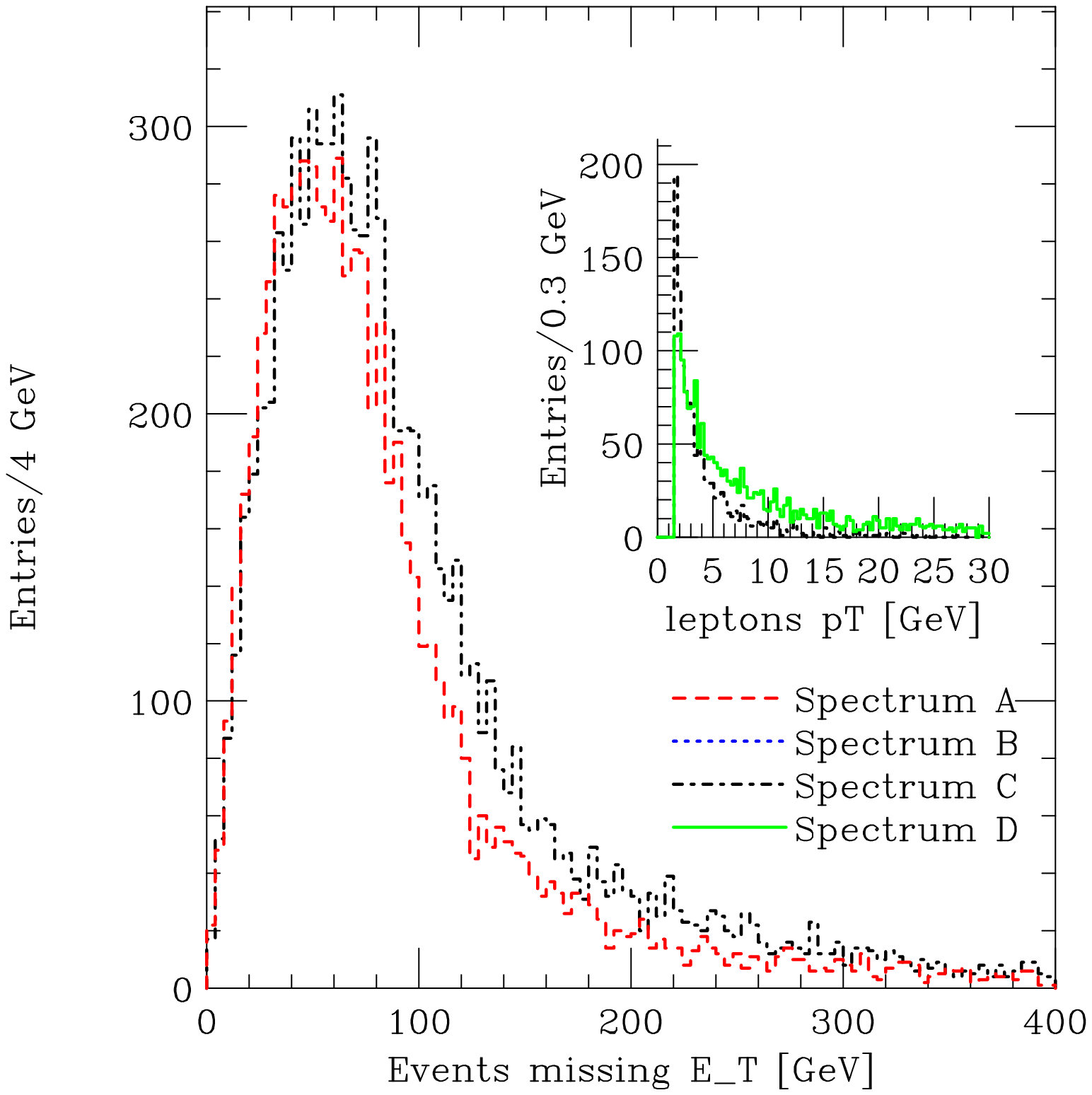} 
  \end{minipage}  
  \end{tabular}
 \caption{This figure shows the missing transverse 
   momentum $p_T$ of leading jets, {\bf plot (a)}, and missing
   $E_T$, {\bf plot (b)}, from generator-level analyses for the example
   pMSSM spectra shown in in Fig.~\ref{fig.spec1}. The leptons $p_T$
   for spectra C and D are also shown on the second plot.} 
 \label{fig2.spec1}
\end{figure}

The $p_T$ and missing $E_T$ features are extracted by performing a
generator-level analysis on SUSY events from 7 TeV LHC simulations for
each of the spectrum A, B, C and D. We use 
\texttt{Herwig++-2.5}~\cite{Gieseke:2011na} to generate 10000
SUSY events from each of the 
spectra points. The four-vectors of particles 
energy and three-momenta are passed on to
\texttt{fastjet-2.4.2}~\cite{Cacciari:2005hq} for obtaining and
sorting the jets from events final states. The anti-$k_T$ jet
clustering algorithm with distance parameter $R=0.4$ is used. The
generator level cuts 
applied are: particle pseudorapidity $|\eta| < 5.0$, and transverse
momentum $p^{part}_T > 1.0$ GeV. Leptons are required to have $|\eta|
< 2.5$ and $p^{lep}_T > 1.5$ GeV, while for jets $p^{jets}_T > 1.5$
GeV. The distributions of leading jets $p_T$ in events with 0-lepton
and two or more jets are shown in Fig.~\ref{fig2.spec1}(a). The
distributions of the events missing $E_T$ and leptons'
$p^{lep}_T $ are shown in Fig.~\ref{fig2.spec1}(b). 

\paragraph*{Triggers/Cuts.}
It is mainly expected 
SUSY collider events 
would end with high-$p_T$ jets and large missing $E_T$. The triggers or
kinematic cuts used in the SUSY search analysis~\cite{daCosta:2011qk}
effectively requires reconstructed jet to have $p_T > 120$ GeV and
events with more that $100$ GeV of missing $p_T$. These cuts would
have thrown away most of scenario A and B events as background events
(see Fig.~\ref{fig2.spec1}) if actually produced during the collider
run. In the absence of discovery from the 2010 
LHC data and the increase in luminosity in current runs the cuts would
be increasing (perhaps to more that 400 GeV on the jets $p_T$ cuts)
in order to be more conservative in events 
selection. As a result MSSM 
spectra which lead to soft jets, soft leptons and/or low, compared to
what is usually assumed, missing transverse momentum would be lost
forever if produced at the LHC. 

\paragraph*{Conclusions.}
The pMSSM posterior samples have MSSM spectra
with several distinct characteristics. A large splitting between the lightest squark and the
LSP is usually assumed as the main characteristic feature of SUSY
spectrum. This 
scenario is indeed very probable as can be seen from
Fig.~\ref{deltM}(a). However, there is a significant region where the
mass difference is small (comparable to the top-quark mass) as in the
peak in $\Delta M_{min}$ around 10 to 25 GeV. A SUSY scenario with the
latter pattern would be difficult to exclude by the LHC experiments
due to the nature of the detectors' trigger system. 
One can conclude that, {\it the LHC with current running parameters
  cannot rule out the MSSM unless  special trigger systems are
  commissioned or if a Higgs boson is found with mass outside the generic MSSM
   prediction.} 
In the case where a Higgs boson is found and with mass in the
range valid for the MSSM points then the discovery or exclusion of
models illustrated here would most likely have to await the operation
of a linear collider~\cite{AguilarSaavedra:2001rg}.   

Thanks to Maurizio Pierini, Umberto De Sanctis, Fernando Quevedo for various
discussions and helpful comments; to Cambridge SUSY working group
for comments and help in fixing some of the software used; and to the 
Instituto de Fesica Teorica (IFT UAM/CSIC), Madrid for hospitality during a visit while work presented were in progress.

\end{document}